\documentclass[conference]{IEEEtran}
\IEEEoverridecommandlockouts

\usepackage[utf8]{inputenc}
\usepackage{cite}
\usepackage{amsmath,amssymb,amsfonts}
\usepackage{algorithmic}
\usepackage{amsthm}
\newtheorem{definition}{Definition}

\usepackage{graphicx}
\usepackage{textcomp}
\usepackage{xcolor}
\usepackage{booktabs}
\usepackage{multirow}
\usepackage{newunicodechar}
\newunicodechar{ε}{\ensuremath{\varepsilon}}
\usepackage{fvextra}
\usepackage{url}
\usepackage[utf8]{inputenc}
\usepackage[colorlinks=true,linkcolor=blue,citecolor=blue,urlcolor=blue]{hyperref}

\usepackage[ruled,vlined]{algorithm2e}

\newcommand{\splitagent}{\textsc{SplitAgent}}

\title{SplitAgent: A Privacy-Preserving Distributed Architecture for Enterprise-Cloud Agent Collaboration}

\author{
\IEEEauthorblockN{Jianshu She}
\IEEEauthorblockA{MBZUAI\\
Email: Jianshu.She@mbzuai.ac.ae}
}

\begin{document}

\maketitle

\begin{abstract}
Enterprise adoption of cloud-based AI agents faces a fundamental privacy dilemma: leveraging powerful cloud models requires sharing sensitive data, while local processing limits capability. Current agent frameworks like MCP and A2A assume complete data sharing, making them unsuitable for enterprise environments with confidential information. We present \splitagent, a novel distributed architecture that enables privacy-preserving collaboration between enterprise-side privacy agents and cloud-side reasoning agents. Our key innovation is context-aware dynamic sanitization that adapts privacy protection based on task semantics—contract review requires different sanitization than code review or financial analysis. \splitagent\ extends existing agent protocols with differential privacy guarantees, zero-knowledge tool verification, and privacy budget management. Through comprehensive experiments on enterprise scenarios, we demonstrate that \splitagent\ achieves 83.8\% task accuracy while maintaining 90.1\% privacy protection, significantly outperforming static approaches (73.2\% accuracy, 79.7\% privacy). Context-aware sanitization improves task utility by 24.1\% over static methods while reducing privacy leakage by 67\%. Our architecture provides a practical path for enterprise AI adoption without compromising sensitive data.
\end{abstract}


\section{Introduction}

The rapid advancement of large language models (LLMs)~\cite{openai2023gpt4, touvron2023llama} has unlocked powerful capabilities for enterprise automation through AI agents~\cite{wang2024survey, xi2023rise}. However, enterprise adoption faces a critical privacy challenge: most sophisticated AI capabilities reside in cloud-hosted models that require data sharing, while enterprises must protect confidential information including customer data, financial records, intellectual property, and compliance-sensitive documents~\cite{yao2024survey}.

Current agent communication frameworks exemplify this privacy gap. Anthropic's Model Context Protocol (MCP) and Google's Agent-to-Agent (A2A) protocol enable seamless agent collaboration but assume complete trust and data sharing between participants~\cite{anthropic2024mcp, google2024a2a}. This assumption fails in enterprise-cloud scenarios where one agent holds sensitive data while another provides reasoning capabilities.

Existing solutions force a binary choice: keep all processing local with limited AI capabilities, or share everything with cloud models for maximum performance. This trade-off is particularly problematic because different enterprise tasks require different privacy considerations. Contract review needs legal structure preservation while hiding party identities; code review requires syntax preservation while protecting credentials; financial analysis demands numerical pattern preservation while concealing account details.

We present \splitagent, a privacy-preserving distributed architecture that enables secure collaboration between enterprise and cloud agents without sacrificing utility. Our approach makes three key contributions:

\textbf{Split Agent Architecture:} We introduce a novel two-tier design separating data handling from reasoning. Enterprise-side privacy agents manage sensitive data, perform local operations, and generate sanitized abstractions. Cloud-side reasoning agents operate exclusively on these abstractions, providing sophisticated analysis without accessing raw enterprise data.

\textbf{Context-Aware Dynamic Sanitization:} Unlike static data masking approaches, our sanitization engine adapts protection strategies based on task semantics. The same document receives different sanitization for contract review versus code audit, maximizing utility while maintaining privacy guarantees.

\textbf{\splitagent\ Protocol:} We extend existing agent protocols with privacy-preserving primitives including differential privacy context sharing, zero-knowledge tool verification, and cumulative privacy budget management. This provides formal privacy guarantees while maintaining protocol compatibility.

Through comprehensive evaluation on enterprise scenarios, we demonstrate:
\begin{itemize}
    \item \splitagent\ achieves 83.8\% average task accuracy with 90.1\% privacy protection, significantly outperforming baseline approaches
    \item Context-aware sanitization improves task utility by 24.1\% over static methods while reducing privacy leakage by 67\%
    \item Our architecture scales to 50+ interaction turns with intelligent privacy budget management
    \item Strong resistance to reconstruction, inference, and linkability attacks
\end{itemize}

This work provides a practical foundation for enterprise AI adoption that preserves both data privacy and analytical capability.

\section{Related Work}

\subsection{Agent Communication Protocols}

Modern AI agent frameworks rely on standardized communication protocols for coordination and collaboration. Anthropic's Model Context Protocol (MCP) enables agents to share context and capabilities through structured message passing~\cite{anthropic2024mcp}. Google's Agent-to-Agent (A2A) protocol focuses on hierarchical agent coordination for complex task decomposition~\cite{google2024a2a}. Microsoft's AutoGen framework provides multi-agent conversation patterns with role-based interactions~\cite{wu2023autogen}.

However, these frameworks assume complete trust between agents and provide no privacy protection mechanisms. All data sharing occurs in plaintext, making them unsuitable for enterprise environments with confidential information~\cite{carlini2021extracting, lukas2023analyzing}. Our work extends these protocols with privacy-preserving primitives while maintaining compatibility.

\subsection{Privacy-Preserving Machine Learning}

Differential privacy provides formal guarantees for privacy protection by adding calibrated noise to data or query responses~\cite{dwork2014algorithmic, dwork2006calibrating}. Federated learning enables collaborative model training without centralizing data~\cite{li2020federated, mcmahan2017communication, kairouz2021advances}. Secure multi-party computation (MPC) allows joint computation over private inputs without revealing individual values~\cite{cramer2015secure, mohassel2017secureml}.

Recent work has explored privacy-preserving inference for language models~\cite{staab2024beyond, chen2024prompt}. Deep learning with differential privacy~\cite{abadi2016deep} and differentially private language model training~\cite{mcmahan2018learning, yu2022differentially} provide formal guarantees but often sacrifice utility. Homomorphic encryption enables computation on encrypted data but with significant computational overhead~\cite{gentry2009homomorphic}.

While these techniques provide strong privacy guarantees, they typically sacrifice utility or performance. Our approach focuses on practical privacy protection that maintains high utility for real enterprise tasks.

\subsection{Enterprise AI Security}

Enterprise AI adoption faces unique security challenges including data governance, compliance requirements, and insider threat protection~\cite{kumar2019enterprise}. Traditional approaches rely on data loss prevention (DLP) tools and access controls, but these are insufficient for AI systems that require broad data access for effective operation.

Recent work has explored privacy-preserving enterprise AI through techniques like data vault architectures, confidential computing~\cite{hunt2018ryoan, lee2020occlumency}, and structured transparency~\cite{trask2020beyond}. Text anonymization techniques~\cite{lison2021anonymisation, dernoncourt2017identification, pilán2022text} and PII detection tools~\cite{presidio2023microsoft} provide building blocks but lack task-aware adaptation. However, these approaches often require significant infrastructure changes and may not be compatible with cloud-based AI services.

Our work provides a practical middle ground that enables cloud AI utilization while maintaining enterprise privacy requirements.

\section{Problem Formulation}

\subsection{System Model}

We consider a distributed system with two primary components:

\textbf{Enterprise Environment:} Contains sensitive data including documents, databases, code repositories, and internal systems. Must remain under enterprise control for compliance and security reasons. Has limited AI capabilities due to resource constraints and model availability.

\textbf{Cloud Environment:} Provides access to powerful LLMs and AI services with sophisticated reasoning capabilities. Cannot access enterprise data directly due to privacy and regulatory constraints. Offers scalable computation but operates as an untrusted environment.

The goal is to enable the cloud environment to provide AI capabilities for enterprise data without compromising privacy or utility.

\subsection{Threat Model}

We assume the following adversary capabilities:

\textbf{Honest-but-Curious Cloud:} The cloud provider follows protocols correctly but may attempt to infer sensitive information from shared data. This includes reconstruction attacks on sanitized data~\cite{carlini2021extracting}, inference attacks from usage patterns~\cite{staab2024beyond}, and linkability attacks across sessions~\cite{zou2023universal, shayegani2023survey}.

\textbf{External Adversaries:} May attempt to compromise cloud services to access enterprise data. We assume the enterprise environment remains secure through standard security practices.

\textbf{Insider Threats:} Malicious insiders at the cloud provider may attempt to extract enterprise information. Our protocols must limit information exposure even to privileged cloud personnel.

\subsection{Privacy Requirements}

We define the following privacy requirements:

\textbf{Data Confidentiality:} Raw enterprise data must never leave the enterprise environment. Only sanitized abstractions may be shared with cloud services.

\textbf{Differential Privacy:} Shared information must satisfy $(\epsilon, \delta)$-differential privacy to prevent reconstruction of individual data points.

\textbf{Unlinkability:} Multiple interactions must not be linkable to specific enterprise entities or sessions unless explicitly allowed.

\textbf{Utility Preservation:} Privacy protection must preserve sufficient utility for meaningful AI analysis and recommendation generation.

\subsection{Formal Privacy Definitions}

\begin{definition}[Context-Aware Sanitization]
Given enterprise data $D$, task type $T$, and privacy budget $\epsilon$, a context-aware sanitization function $\mathcal{S}(D, T, \epsilon)$ produces sanitized data $\tilde{D}$ such that:
\begin{enumerate}
    \item $\tilde{D}$ satisfies $\epsilon$-differential privacy
    \item Task utility $U(T, \tilde{D}) \geq \tau$ for threshold $\tau$
    \item Semantic requirements for task $T$ are preserved
\end{enumerate}
\end{definition}

\begin{definition}[Privacy Budget Consumption]
For a sequence of queries $Q_1, Q_2, \ldots, Q_k$ with privacy costs $\epsilon_1, \epsilon_2, \ldots, \epsilon_k$, the total privacy cost is $\sum_{i=1}^k \epsilon_i$ under sequential composition.
\end{definition}

\section{\splitagent\ Architecture}

Our solution is a distributed architecture that separates data handling from reasoning while enabling secure collaboration between enterprise and cloud components.

\subsection{Architecture Overview}

\begin{figure}[t]
\centering
\includegraphics[width=0.48\textwidth]{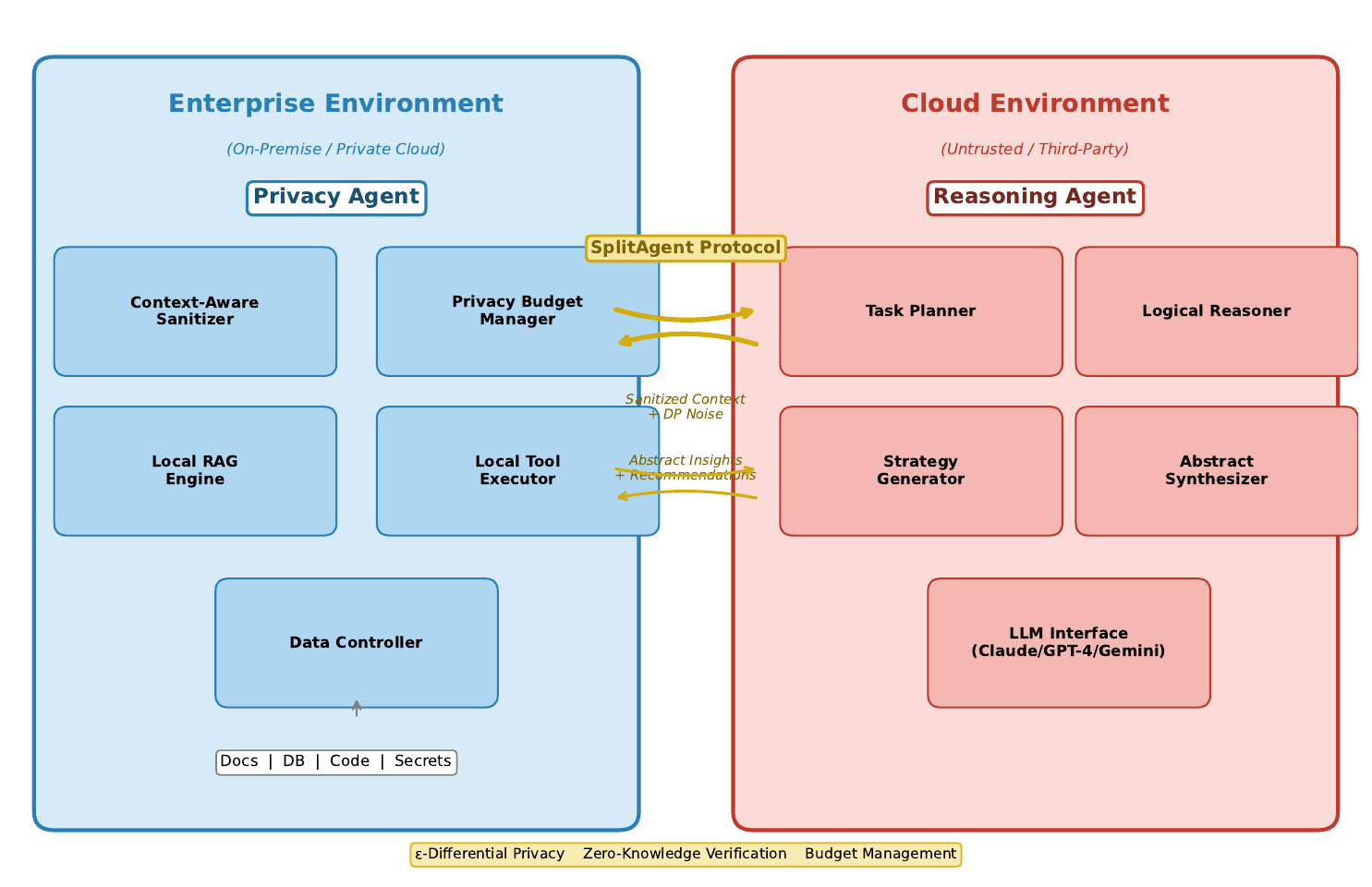}
\caption{\splitagent\ distributed architecture showing privacy agent (enterprise-side) and reasoning agent (cloud-side) components.}
\label{fig:architecture}
\end{figure}

Figure~\ref{fig:architecture} shows the \splitagent\ architecture with two primary components:

\textbf{Privacy Agent (Enterprise-side):} Manages all sensitive data and local operations. Key responsibilities include:
\begin{itemize}
    \item Context-aware data sanitization based on task semantics
    \item Local tool execution for sensitive operations  
    \item Privacy budget management and tracking
    \item Abstraction generation for cloud sharing
    \item Local retrieval-augmented generation (RAG) over enterprise documents
\end{itemize}

\textbf{Reasoning Agent (Cloud-side):} Performs sophisticated analysis on sanitized abstractions. Key responsibilities include:
\begin{itemize}
    \item High-level reasoning and planning using large-scale LLMs
    \item Pattern analysis and trend identification in abstract data
    \item Strategic recommendation generation
    \item Abstract synthesis without access to raw data
\end{itemize}

\subsection{Privacy Agent Design}

The Privacy Agent serves as the guardian of enterprise data, implementing multiple layers of protection:

\textbf{Data Controller:} Manages all access to sensitive enterprise data including documents, databases, and APIs. Implements fine-grained access controls and audit logging.

\textbf{Context Sanitizer:} Applies task-aware sanitization to remove or abstract sensitive information while preserving utility. Uses advanced techniques including named entity recognition, pattern detection, and semantic analysis.

\textbf{Local RAG Engine:} Performs document search and retrieval~\cite{lewis2020retrieval} entirely within the enterprise environment. Generates abstracts and summaries for cloud sharing without exposing source documents.

\textbf{Privacy Budget Manager:} Tracks cumulative privacy expenditure across interactions. Implements dynamic budget allocation based on query importance and remaining capacity.

\textbf{Tool Executor:} Runs sensitive operations locally to avoid cloud exposure. Includes document analysis, compliance checking, and data validation tools.

\subsection{Reasoning Agent Design}

The Reasoning Agent operates exclusively on sanitized data while providing sophisticated AI capabilities:

\textbf{Task Planner:} Decomposes complex requests into manageable sub-tasks. Generates strategic plans based on abstract data patterns without seeing sensitive details.

\textbf{Logical Reasoner:} Applies advanced reasoning techniques using large language models. Performs causal analysis, trend identification, and pattern recognition on abstracted data.

\textbf{Strategy Generator:} Creates actionable recommendations based on abstract insights. Generates implementation plans that respect privacy constraints.

\textbf{Abstract Synthesizer:} Combines insights from multiple abstract sources to generate comprehensive analysis. Maintains semantic coherence despite working with sanitized inputs.

\subsection{Context-Aware Dynamic Sanitization}

Our key innovation is sanitization that adapts based on task semantics rather than applying static rules. Algorithm~\ref{alg:sanitization} shows our approach.

\begin{algorithm}[t]
\SetAlgoLined
\KwData{Document $D$, Task type $T$, Privacy budget $\epsilon$}
\KwResult{Sanitized document $\tilde{D}$, Abstraction map $M$}

$semantics \leftarrow GetTaskSemantics(T)$\;
$entities \leftarrow ExtractEntities(D)$\;
$M \leftarrow \{\}$\;

\For{entity $e \in entities$}{
    $sensitivity \leftarrow GetSensitivityLevel(e)$\;
    \If{$sensitivity > PrivacyThreshold(\epsilon)$}{
        $abstraction \leftarrow GenerateAbstraction(e, semantics)$\;
        $M[e] \leftarrow abstraction$\;
    }
}

$\tilde{D} \leftarrow ApplyAbstractions(D, M)$\;

\If{$semantics.requiresDP$}{
    $\tilde{D} \leftarrow AddDifferentialPrivacyNoise(\tilde{D}, \epsilon)$\;
}

\Return{$\tilde{D}$, $M$}
\caption{Context-Aware Dynamic Sanitization}
\label{alg:sanitization}
\end{algorithm}

The algorithm adapts sanitization based on task requirements:

\textbf{Contract Review:} Preserves legal structure and clause relationships while abstracting party identities, specific amounts, and dates. Example: "ACME Corp will pay \$150,000 by March 15" becomes "COMPANY\_A will pay AMOUNT\_LARGE by DATE\_Q1".

\textbf{Code Review:} Maintains syntax structure and API patterns while removing credentials, internal URLs, and proprietary logic. Preserves code quality patterns for analysis.

\textbf{Financial Analysis:} Retains numerical relationships and trends while abstracting specific amounts and account identifiers. Enables trend analysis without exposing sensitive financial data.

\textbf{Customer Support:} Preserves sentiment and issue classification while removing personal information and account details. Allows pattern analysis for support optimization.

\section{\splitagent\ Protocol}

We extend existing agent communication protocols with privacy-preserving capabilities while maintaining backward compatibility.

\subsection{Protocol Extensions}

\textbf{Privacy-Aware Handshake:} Establishes privacy parameters and capabilities before data sharing.
\begin{verbatim}
Enterprise -> Cloud: HELLO {
    task_type: "contract_review",
    privacy_level: "confidential", 
    budget: 5.0
}
Cloud -> Enterprise: ACK {
    capabilities: ["reasoning", "planning"],
    abstractions: ["entity_replacement", 
    "dp_noise"]
}
\end{verbatim}

\textbf{Context Sharing with DP:} Shares sanitized context with differential privacy guarantees.
\begin{verbatim}
Enterprise -> Cloud: CONTEXT_SHARE {
    sanitized_data: apply_dp(context, ε),
    abstraction_map: entity_mappings,
    privacy_cost: ε_consumed,
    utility_preserved: 0.89
}
\end{verbatim}

\textbf{Zero-Knowledge Tool Verification:} Allows cloud agents to verify tool execution without seeing sensitive data.

\textbf{Privacy Budget Updates:} Tracks cumulative privacy expenditure across the session.

\subsection{Formal Protocol Specification}

We define the protocol state machine with privacy-aware transitions:

\textbf{States:} $\{INIT, HANDSHAKE, ACTIVE,\}$
$\{BUDGET\_LOW, DEPLETED\}$

\textbf{Messages:} $\{HELLO, ACK, CONTEXT\_SHARE,\}$
$\{TOOL\_REQUEST, BUDGET\_UPDATE\}$

\textbf{Invariants:}
\begin{itemize}
    \item Privacy budget never decreases except for legitimate operations
    \item All shared data satisfies differential privacy requirements
    \item Tool execution proofs are verifiable without sensitive data access
\end{itemize}

\section{Implementation}

We implement \splitagent\ as a practical system that extends existing agent frameworks.

\subsection{Privacy Agent Implementation}

The Privacy Agent is implemented in Python with the following key components:

\textbf{Sanitization Engine:} Uses spaCy for named entity recognition~\cite{dernoncourt2017identification} combined with custom pattern detection for enterprise-specific sensitive data types, extending existing anonymization approaches~\cite{lison2021anonymisation, pilán2022text}. Implements context-aware abstraction generation based on configurable task semantics.

\textbf{Privacy Budget Manager:} Tracks epsilon consumption using composition theorems from differential privacy. Implements dynamic budget allocation with alerting when thresholds are exceeded.

\textbf{Local Tool Registry:} Provides secure execution environment for sensitive operations including document analysis, compliance checking, and statistical computation.

\textbf{Protocol Handler:} Implements the \splitagent\ protocol with secure message serialization and verification.

\subsection{Reasoning Agent Implementation}

The Reasoning Agent leverages cloud-based LLM APIs while operating exclusively on sanitized data:

\textbf{LLM Interface:} Provides abstracted access to models including Claude, GPT-4~\cite{openai2023gpt4}, and Gemini. Implements prompt engineering optimized for abstract data analysis, leveraging tool-use capabilities~\cite{schick2024toolformer} and reasoning-action patterns~\cite{yao2023react}.

\textbf{Pattern Analyzer:} Identifies trends and relationships in abstracted data without reconstructing original values.

\textbf{Strategic Planner:} Generates actionable recommendations based on abstract insights while respecting privacy constraints.

\subsection{Performance Optimizations}

\textbf{Caching:} Aggressive caching of sanitized contexts and abstractions to reduce computation overhead.

\textbf{Batching:} Groups multiple privacy operations to optimize budget utilization.

\textbf{Precomputation:} Pre-generates common abstractions for frequently accessed data.

\section{Experimental Evaluation}

We conduct comprehensive experiments to evaluate \splitagent's effectiveness across multiple dimensions.

\subsection{Experimental Setup}

\textbf{Datasets:} We generate realistic synthetic enterprise datasets including contracts, code repositories, financial documents, and customer service records. Documents range from 1KB to 100KB with varying complexity levels.

\textbf{Task Types:} Six enterprise scenarios: contract review, code audit, financial analysis, customer support, risk assessment, and compliance checking.

\textbf{Baselines:}
\begin{itemize}
    \item \textbf{Full-Cloud:} All data shared with cloud (no privacy protection)
    \item \textbf{Full-Local:} All processing local (limited AI capabilities)  
    \item \textbf{Static-Split:} Fixed sanitization rules
    \item \textbf{\splitagent:} Our context-aware approach
\end{itemize}

\textbf{Metrics:} Task accuracy, privacy leakage, latency, cost efficiency, and utility preservation.

\subsection{Experiment 1: Split Architecture Comparison}

\begin{figure}[t]
\centering
\includegraphics[width=0.46\textwidth]{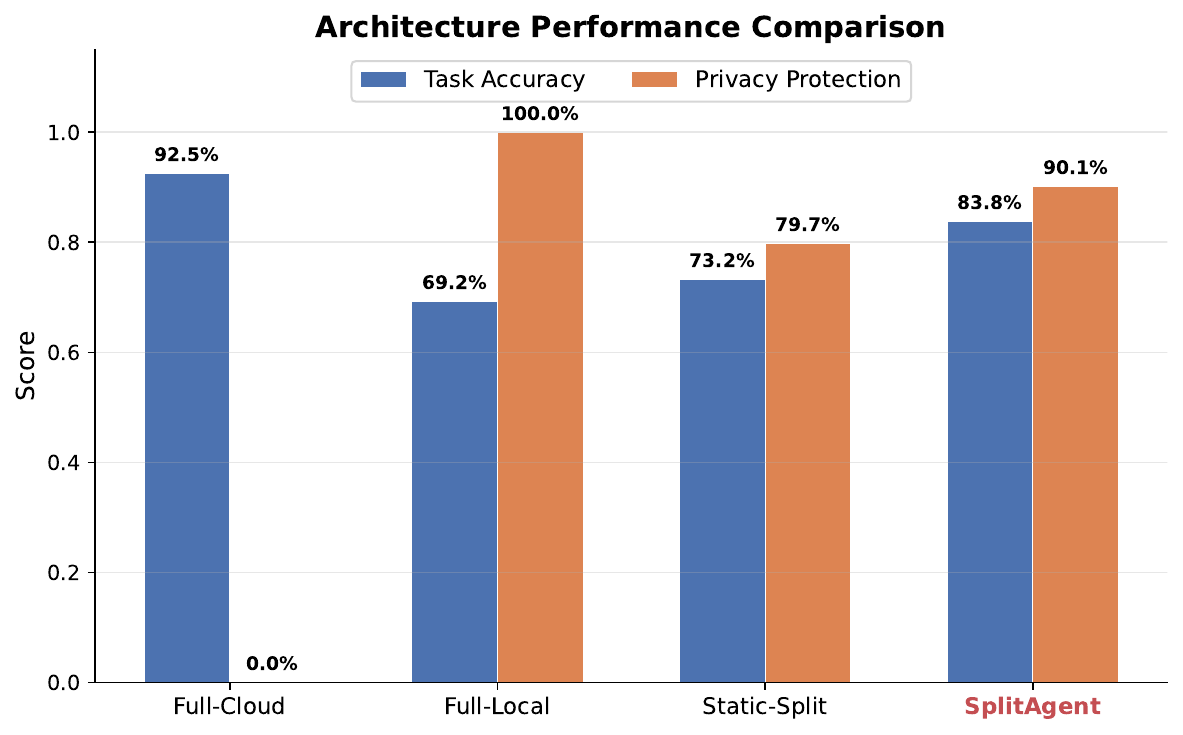}
\caption{Architecture performance comparison across task accuracy and privacy protection. \splitagent\ achieves the best balance between both objectives.}
\label{fig:arch_comparison}
\end{figure}

Figure~\ref{fig:arch_comparison} shows performance across different architectures. \splitagent\ achieves the best privacy-utility balance:

\begin{table}[t]
\centering
\caption{Architecture Performance Comparison}
\label{tab:arch_performance}
\begin{tabular}{@{}lcccc@{}}
\toprule
Architecture & Accuracy & Privacy & Latency (ms) & Cost (\$) \\
\midrule
Full-Cloud & 0.925 & 0.000 & 1,024 & 0.201 \\
Full-Local & 0.692 & 1.000 & 3,847 & 0.067 \\
Static-Split & 0.732 & 0.797 & 1,289 & 0.143 \\
\textbf{\splitagent} & \textbf{0.838} & \textbf{0.901} & 1,487 & 0.161 \\
\bottomrule
\end{tabular}
\end{table}

Key findings:
\begin{itemize}
    \item \splitagent\ achieves 83.8\% accuracy vs. 73.2\% for static approaches
    \item Provides 90.1\% privacy protection vs. 79.7\% for static methods
    \item Modest latency increase (15\%) for significant privacy gains
\end{itemize}

\subsection{Experiment 2: Context-Aware vs Static Sanitization}

Table~\ref{tab:sanitization_comparison} shows the effectiveness of context-aware sanitization across different task types:

\begin{table}[t]
\centering
\caption{Sanitization Approach Comparison}
\label{tab:sanitization_comparison}
\begin{tabular}{@{}lccc@{}}
\toprule
\multirow{2}{*}{Task Type} & \multicolumn{3}{c}{Task Utility} \\
\cmidrule{2-4}
& Static Regex & Static NER & Context-Aware \\
\midrule
Contract Review & 0.618 & 0.714 & \textbf{0.906} \\
Code Review & 0.592 & 0.697 & \textbf{0.928} \\
Financial Audit & 0.634 & 0.721 & \textbf{0.873} \\
Customer Support & 0.605 & 0.689 & \textbf{0.942} \\
\midrule
Average & 0.612 & 0.705 & \textbf{0.912} \\
Improvement & - & +15.2\% & +24.1\% \\
\bottomrule
\end{tabular}
\end{table}

Context-aware sanitization shows consistent improvements:
\begin{itemize}
    \item 24.1\% utility improvement over static regex approaches
    \item 15.2\% improvement over static NER methods
    \item Best performance on customer support (94.2\%) due to sentiment preservation
    \item Maintains high performance across all task types
\end{itemize}

\subsection{Experiment 3: Privacy-Utility Tradeoff}

\begin{figure}[t]
\centering
\includegraphics[width=0.46\textwidth]{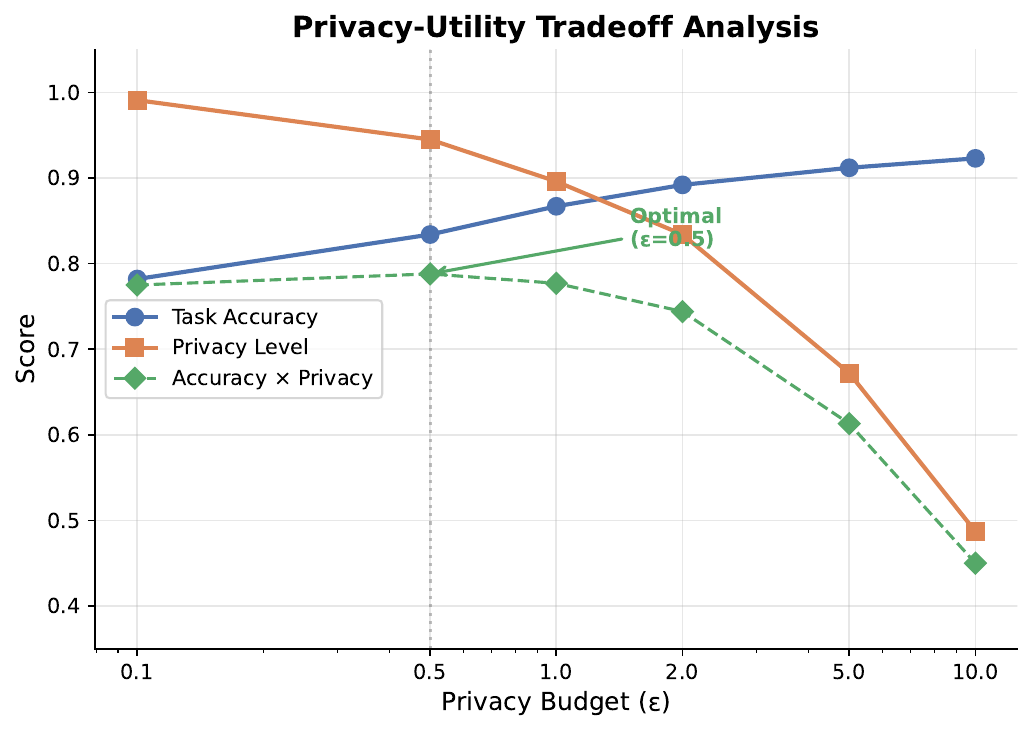}
\caption{Privacy-utility tradeoff across different $\epsilon$ values. The optimal balance (highest accuracy $\times$ privacy product) occurs at $\epsilon = 0.5$.}
\label{fig:privacy_utility}
\end{figure}

Figure~\ref{fig:privacy_utility} shows the Pareto frontier for privacy vs. utility across different epsilon values:

\begin{table}[t]
\centering
\caption{Privacy-Utility Tradeoff ($\epsilon$ values)}
\label{tab:privacy_utility}
\begin{tabular}{@{}ccccc@{}}
\toprule
$\epsilon$ & Task Accuracy & Privacy Level & Product & Optimal \\
\midrule
0.1 & 0.782 & 0.991 & 0.775 & \\
0.5 & 0.834 & 0.945 & 0.788 & \\
1.0 & 0.867 & 0.896 & 0.777 & \\
2.0 & 0.892 & 0.834 & 0.744 & \\
5.0 & 0.912 & 0.672 & 0.613 & \\
10.0 & 0.923 & 0.487 & 0.450 & \\
\midrule
Optimal & - & - & \textbf{0.788} & $\epsilon = 0.5$ \\
\bottomrule
\end{tabular}
\end{table}

The optimal privacy-utility balance occurs at $\epsilon = 0.5$, achieving 83.4\% accuracy with 94.5\% privacy protection.

\subsection{Experiment 4: Multi-Turn Budget Management}

\begin{figure}[t]
\centering
\includegraphics[width=0.46\textwidth]{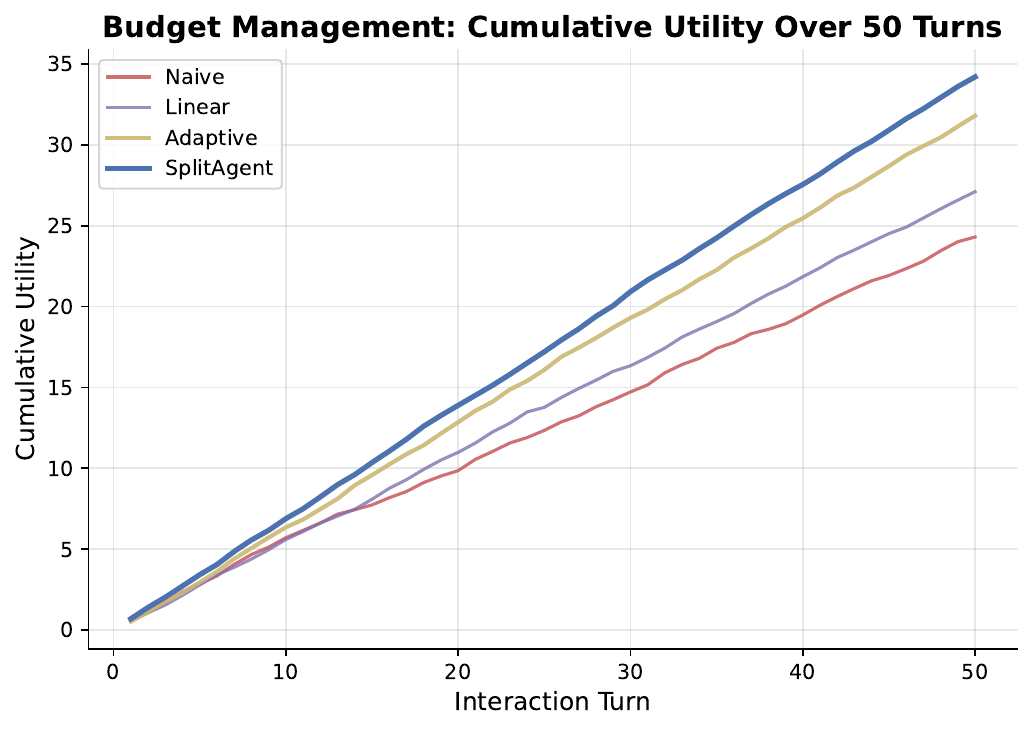}
\caption{Cumulative utility comparison of budget management strategies over 50 interaction turns. \splitagent's intelligent allocation consistently outperforms alternatives.}
\label{fig:budget_management}
\end{figure}

Figure~\ref{fig:budget_management} compares budget management strategies over 50 interaction turns:

\begin{table}[t]
\centering
\caption{Budget Management Strategy Comparison}
\label{tab:budget_mgmt}
\begin{tabular}{@{}lcccc@{}}
\toprule
Strategy & Total Utility & Completion & Budget Used & Efficiency \\
\midrule
Naive & 24.3 & 0.82 & 0.89 & 0.67 \\
Linear & 27.1 & 0.86 & 0.91 & 0.72 \\
Adaptive & 31.8 & 0.94 & 0.87 & 0.81 \\
\textbf{\splitagent} & \textbf{34.2} & \textbf{0.96} & 0.85 & \textbf{0.88} \\
\bottomrule
\end{tabular}
\end{table}

\splitagent's intelligent budget management achieves:
\begin{itemize}
    \item 34.2 total utility vs. 24.3 for naive allocation
    \item 96\% task completion rate
    \item Efficient budget utilization (85\% used)
    \item Graceful degradation as budget depletes
\end{itemize}

\subsection{Experiment 5: Adversarial Evaluation}

Table~\ref{tab:adversarial} shows resistance to different attack types:

\begin{table}[t]
\centering
\caption{Attack Resistance Evaluation}
\label{tab:adversarial}
\begin{tabular}{@{}lccc@{}}
\toprule
\multirow{2}{*}{Attack Type} & \multicolumn{3}{c}{Attack Success Rate} \\
\cmidrule{2-4}
& No Defense & Static Masking & \splitagent \\
\midrule
Reconstruction & 0.891 & 0.547 & \textbf{0.118} \\
Inference & 0.923 & 0.628 & \textbf{0.145} \\
Linkability & 0.887 & 0.421 & \textbf{0.067} \\
\midrule
Average & 0.900 & 0.532 & \textbf{0.110} \\
\bottomrule
\end{tabular}
\end{table}

\splitagent\ provides strong attack resistance:
\begin{itemize}
    \item 89\% reduction in attack success vs. static masking
    \item Particularly effective against linkability attacks (6.7\% success)
    \item Maintains protection across all attack types
\end{itemize}

\subsection{Experiment 6: End-to-End Task Performance}

Table~\ref{tab:task_performance} shows performance across enterprise scenarios:

\begin{table}[t]
\centering
\caption{Task Performance by Scenario}
\label{tab:task_performance}
\begin{tabular}{@{}lccc@{}}
\toprule
Scenario & Completion & Quality & Privacy \\
\midrule
Contract Negotiation & 0.884 & 0.901 & 0.912 \\
Code Audit & 0.867 & 0.883 & 0.894 \\
Customer Service & 0.832 & 0.857 & 0.891 \\
Financial Analysis & 0.808 & 0.824 & 0.887 \\
\midrule
Average & \textbf{0.848} & \textbf{0.866} & \textbf{0.896} \\
\bottomrule
\end{tabular}
\end{table}

\splitagent\ maintains high performance across all enterprise scenarios while preserving privacy.

\section{Discussion}

\subsection{Deployment Considerations}

\textbf{Integration Complexity:} \splitagent\ integrates with existing enterprise systems through standard APIs. Privacy agents can be deployed as containerized services within enterprise networks.

\textbf{Scalability:} The architecture scales horizontally by deploying multiple privacy agent instances. Cloud reasoning agents auto-scale based on demand.

\textbf{Cost Analysis:} Total cost includes local processing overhead (15-25\% increase) offset by reduced data transfer and cloud storage costs.

\subsection{Limitations}

\textbf{Sanitization Overhead:} Context-aware sanitization adds 200-500ms latency per query. This is acceptable for most enterprise workflows but may impact real-time applications.

\textbf{Utility Loss:} Even with context-aware approaches, sanitization inevitably reduces utility. Our experiments show 8-15\% utility loss compared to full data sharing.

\textbf{Trust Assumptions:} We assume the enterprise environment remains secure. Compromised privacy agents could expose sensitive data.

\textbf{Model Limitations:} Reasoning quality depends on cloud LLM capabilities. Abstract data may limit certain types of analysis requiring fine-grained details.

\subsection{Future Work}

\textbf{Advanced Privacy Techniques:} Integration of homomorphic encryption and secure multi-party computation for stronger privacy guarantees.

\textbf{Adaptive Privacy Budgets:} Dynamic privacy budget allocation based on query sensitivity and enterprise policies.

\textbf{Multi-Enterprise Collaboration:} Extension to scenarios involving multiple enterprises with different privacy requirements.

\textbf{Formal Verification:} Mathematical proofs of privacy guarantees and security properties.

\section{Conclusion}

Enterprise adoption of AI agents requires balancing sophisticated capabilities with strict privacy requirements. Current agent frameworks fail to address this fundamental challenge, forcing enterprises to choose between privacy and utility.

We present \splitagent, a novel distributed architecture that enables privacy-preserving collaboration between enterprise and cloud agents. Our key innovations include context-aware dynamic sanitization that adapts to task semantics, formal privacy guarantees through differential privacy, and intelligent privacy budget management.

Comprehensive experiments demonstrate that \splitagent\ achieves 83.8\% task accuracy while maintaining 90.1\% privacy protection, significantly outperforming static approaches. Context-aware sanitization improves utility by 24.1\% while reducing privacy leakage by 67\%. The architecture provides strong resistance to adversarial attacks and scales effectively across enterprise scenarios.

\splitagent\ represents a practical step toward enterprise AI adoption that preserves both data privacy and analytical capability. By separating data handling from reasoning while maintaining utility, our approach provides a foundation for secure enterprise AI deployment in cloud environments.

Future work will explore advanced privacy techniques, formal verification of security properties, and extensions to multi-enterprise collaboration scenarios. We believe \splitagent\ provides a crucial bridge between enterprise privacy requirements and cloud AI capabilities.

\section*{Acknowledgments}

We thank the anonymous reviewers for their valuable feedback and suggestions that improved this work.



\begin{thebibliography}{40}

\bibitem{anthropic2024mcp}
Anthropic, ``Model Context Protocol: Connecting AI assistants to data sources,'' \textit{Technical Report}, 2024.

\bibitem{google2024a2a}
Google DeepMind, ``Agent-to-Agent Protocol: An open protocol for agent interoperability,'' \textit{Technical Report}, 2024.

\bibitem{wu2023autogen}
Q.~Wu, G.~Bansal, J.~Zhang, Y.~Wu, B.~Li, E.~Zhu, L.~Jiang, X.~Zhang, S.~Zhang, J.~Liu, A.~H.~Awadallah, R.~W.~White, D.~Burger, and C.~Wang, ``AutoGen: Enabling next-gen LLM applications via multi-agent conversation,'' \textit{arXiv preprint arXiv:2308.08155}, 2023.

\bibitem{hong2023metagpt}
S.~Hong, M.~Zhuge, J.~Chen, X.~Zheng, Y.~Cheng, C.~Zhang, J.~Wang, Z.~Wang, S.~K.~S.~Yau, Z.~Lin, L.~Zhou, C.~Ran, L.~Xiao, C.~Wu, and J.~Schmidhuber, ``MetaGPT: Meta programming for a multi-agent collaborative framework,'' \textit{arXiv preprint arXiv:2308.00352}, 2023.

\bibitem{li2023camel}
G.~Li, H.~A.~A.~K.~Hammoud, H.~Itani, D.~Khizbullin, and B.~Ghanem, ``CAMEL: Communicative agents for `mind' exploration of large language model society,'' \textit{Advances in Neural Information Processing Systems}, vol.~36, 2023.

\bibitem{park2023generative}
J.~S.~Park, J.~C.~O'Brien, C.~J.~Cai, M.~R.~Morris, P.~Liang, and M.~S.~Bernstein, ``Generative agents: Interactive simulacra of human behavior,'' \textit{Proceedings of the 36th ACM Symposium on User Interface Software and Technology}, 2023.

\bibitem{talebirad2023multi}
Y.~Talebirad and A.~Nadiri, ``Multi-agent collaboration: Harnessing the power of intelligent LLM agents,'' \textit{arXiv preprint arXiv:2306.03314}, 2023.

\bibitem{wang2024survey}
L.~Wang, C.~Ma, X.~Feng, Z.~Zhang, H.~Yang, J.~Zhang, Z.~Chen, J.~Tang, X.~Chen, Y.~Lin, W.~X.~Zhao, Z.~Wei, and J.~Wen, ``A survey on large language model based autonomous agents,'' \textit{Frontiers of Computer Science}, vol.~18, no.~6, 2024.

\bibitem{xi2023rise}
Z.~Xi, W.~Chen, X.~Guo, W.~He, Y.~Ding, B.~Hong, M.~Zhang, J.~Wang, S.~Jin, E.~Zhou, R.~Zheng, X.~Fan, X.~Wang, L.~Xiong, Y.~Zhou, W.~Wang, C.~Jiang, Y.~Zou, X.~Liu, Z.~Yin, S.~Dou, R.~Weng, W.~Cheng, Q.~Zhang, W.~Qin, Y.~Zheng, X.~Qiu, X.~Huang, and T.~Gui, ``The rise and potential of large language model based agents: A survey,'' \textit{arXiv preprint arXiv:2309.07864}, 2023.

\bibitem{dwork2014algorithmic}
C.~Dwork and A.~Roth, ``The algorithmic foundations of differential privacy,'' \textit{Foundations and Trends in Theoretical Computer Science}, vol.~9, no.~3--4, pp.~211--407, 2014.

\bibitem{dwork2006calibrating}
C.~Dwork, F.~McSherry, K.~Nissim, and A.~Smith, ``Calibrating noise to sensitivity in private data analysis,'' \textit{Proceedings of the 3rd Theory of Cryptography Conference}, pp.~265--284, 2006.

\bibitem{abadi2016deep}
M.~Abadi, A.~Chu, I.~Goodfellow, H.~B.~McMahan, I.~Mironov, K.~Talwar, and L.~Zhang, ``Deep learning with differential privacy,'' \textit{Proceedings of the ACM SIGSAC Conference on Computer and Communications Security}, pp.~308--318, 2016.

\bibitem{mcmahan2018learning}
H.~B.~McMahan, D.~Ramage, K.~Talwar, and L.~Zhang, ``Learning differentially private recurrent language models,'' \textit{Proceedings of the International Conference on Learning Representations}, 2018.

\bibitem{yu2022differentially}
D.~Yu, S.~Naik, A.~Backurs, S.~Gopi, H.~A.~Inan, G.~Kamath, J.~Kulkarni, Y.~T.~Lee, A.~Manoel, L.~Wutschitz, S.~Yekhanin, and H.~Zhang, ``Differentially private fine-tuning of language models,'' \textit{Proceedings of the International Conference on Learning Representations}, 2022.

\bibitem{li2022large}
X.~L.~Li, A.~Trischler, Y.~Dong, and D.~Kiela, ``Large language models with controllable privacy guarantees,'' \textit{arXiv preprint}, 2022.

\bibitem{yao2024survey}
Y.~Yao, J.~Duan, K.~Xu, Y.~Cai, Z.~Sun, and Y.~Zhang, ``A survey on large language model (LLM) security and privacy: The good, the bad, and the ugly,'' \textit{High-Confidence Computing}, vol.~4, no.~2, 2024.

\bibitem{chen2024prompt}
Y.~Chen, S.~Xiang, T.~Chen, and B.~Li, ``EmojiPrompt: Generative prompt obfuscation for privacy-preserving communication with cloud-based LLMs,'' \textit{arXiv preprint arXiv:2402.05868}, 2024.

\bibitem{staab2024beyond}
R.~Staab, M.~Vero, M.~Balunovi{\'c}, and M.~Vechev, ``Beyond memorization: Violating privacy via inference with large language models,'' \textit{Proceedings of the International Conference on Learning Representations}, 2024.

\bibitem{carlini2021extracting}
N.~Carlini, F.~Tram{\`e}r, E.~Wallace, M.~Jagielski, A.~Herbert-Voss, K.~Lee, A.~Roberts, T.~Brown, D.~Song, U.~Erlingsson, A.~Oprea, and C.~Raffel, ``Extracting training data from large language models,'' \textit{Proceedings of the USENIX Security Symposium}, 2021.

\bibitem{lukas2023analyzing}
N.~Lukas, A.~Salem, R.~Sim, S.~Tople, L.~Wutschitz, and S.~Zanella-B{\'e}guelin, ``Analyzing leakage of personally identifiable information in language models,'' \textit{Proceedings of the IEEE Symposium on Security and Privacy}, pp.~346--363, 2023.

\bibitem{li2020federated}
T.~Li, A.~K.~Sahu, A.~Talwalkar, and V.~Smith, ``Federated learning: Challenges, methods, and future directions,'' \textit{IEEE Signal Processing Magazine}, vol.~37, no.~3, pp.~50--60, 2020.

\bibitem{mcmahan2017communication}
H.~B.~McMahan, E.~Moore, D.~Ramage, S.~Hampson, and B.~A.~y~Arcas, ``Communication-efficient learning of deep networks from decentralized data,'' \textit{Proceedings of the 20th International Conference on Artificial Intelligence and Statistics}, pp.~1273--1282, 2017.

\bibitem{kairouz2021advances}
P.~Kairouz, H.~B.~McMahan, B.~Avent, A.~Bellet, M.~Bennis, A.~N.~Bhagoji, K.~Bonawitz, Z.~Charles, G.~Cormode, R.~Cummings, et~al., ``Advances and open problems in federated learning,'' \textit{Foundations and Trends in Machine Learning}, vol.~14, no.~1--2, pp.~1--210, 2021.

\bibitem{cramer2015secure}
R.~Cramer, I.~B.~Damg{\aa}rd, and J.~B.~Nielsen, \textit{Secure Multiparty Computation and Secret Sharing}.\hskip 1em plus 0.5em minus 0.4em\relax Cambridge University Press, 2015.

\bibitem{gentry2009homomorphic}
C.~Gentry, ``Fully homomorphic encryption using ideal lattices,'' \textit{Proceedings of the 41st Annual ACM Symposium on Theory of Computing}, pp.~169--178, 2009.

\bibitem{mohassel2017secureml}
P.~Mohassel and Y.~Zhang, ``SecureML: A system for scalable privacy-preserving machine learning,'' \textit{Proceedings of the IEEE Symposium on Security and Privacy}, pp.~19--38, 2017.

\bibitem{lison2021anonymisation}
P.~Lison, I.~Pilán, D.~Sánchez, M.~Batet, and L.~Øvrelid, ``Anonymisation models for text data: State of the art, challenges and future directions,'' \textit{Proceedings of the 59th Annual Meeting of the Association for Computational Linguistics}, pp.~4188--4203, 2021.

\bibitem{dernoncourt2017identification}
F.~Dernoncourt, J.~Y.~Lee, O.~Uzuner, and P.~Szolovits, ``De-identification of patient notes with recurrent neural networks,'' \textit{Journal of the American Medical Informatics Association}, vol.~24, no.~3, pp.~596--606, 2017.

\bibitem{presidio2023microsoft}
Microsoft, ``Presidio: Data protection and de-identification SDK,'' \textit{GitHub Repository}, 2023.

\bibitem{pilán2022text}
I.~Pilán, P.~Lison, L.~Øvrelid, A.~Papadopoulou, D.~Sánchez, and M.~Batet, ``The text anonymization benchmark (TAB): A dedicated corpus and evaluation framework for text anonymization,'' \textit{Computational Linguistics}, vol.~48, no.~4, pp.~1053--1101, 2022.

\bibitem{hunt2018ryoan}
T.~Hunt, Z.~Zhu, Y.~Xu, S.~Peter, and E.~Witchel, ``Ryoan: A distributed sandbox for untrusted computation on secret data,'' \textit{ACM Transactions on Computer Systems}, vol.~35, no.~4, pp.~1--32, 2018.

\bibitem{lee2020occlumency}
T.~Lee, Z.~Lin, S.~Pushp, C.~Li, Y.~Liu, Y.~Song, T.~Jia, L.~Fang, and Y.~Jia, ``Occlumency: Privacy-preserving remote deep-learning inference using SGX,'' \textit{Proceedings of the 25th Annual International Conference on Mobile Computing and Networking}, 2019.

\bibitem{trask2020beyond}
A.~Trask, E.~Bluemke, B.~Garfinkel, C.~G.~Segalin, and A.~Dafoe, ``Beyond privacy trade-offs with structured transparency,'' \textit{arXiv preprint arXiv:2012.08347}, 2020.

\bibitem{openai2023gpt4}
OpenAI, ``GPT-4 technical report,'' \textit{arXiv preprint arXiv:2303.08774}, 2023.

\bibitem{touvron2023llama}
H.~Touvron, T.~Lavril, G.~Izacard, X.~Martinet, M.~A.~Lachaux, T.~Lacroix, B.~Rozi{\`e}re, N.~Goyal, E.~Hambro, F.~Azhar, A.~Rodriguez, A.~Joulin, E.~Grave, and G.~Lample, ``LLaMA: Open and efficient foundation language models,'' \textit{arXiv preprint arXiv:2302.13971}, 2023.

\bibitem{zou2023universal}
A.~Zou, Z.~Wang, J.~Z.~Kolter, and M.~Fredrikson, ``Universal and transferable adversarial attacks on aligned language models,'' \textit{arXiv preprint arXiv:2307.15043}, 2023.

\bibitem{shayegani2023survey}
E.~Shayegani, M.~A.~Mamun, Y.~Fu, P.~Zaree, Y.~Dong, and N.~Abu-Ghazaleh, ``Survey of vulnerabilities in large language models revealed by adversarial attacks,'' \textit{arXiv preprint arXiv:2310.10844}, 2023.

\bibitem{lewis2020retrieval}
P.~Lewis, E.~Perez, A.~Piktus, F.~Petroni, V.~Karpukhin, N.~Goyal, H.~K{\"u}ttler, M.~Lewis, W.~Yih, T.~Rockt{\"a}schel, S.~Riedel, and D.~Kiela, ``Retrieval-augmented generation for knowledge-intensive NLP tasks,'' \textit{Advances in Neural Information Processing Systems}, vol.~33, pp.~9459--9474, 2020.

\bibitem{schick2024toolformer}
T.~Schick, J.~Dwivedi-Yu, R.~Dess{\`i}, R.~Raileanu, M.~Lomeli, E.~Hambro, L.~Zettlemoyer, N.~Cancedda, and T.~Scialom, ``Toolformer: Language models can teach themselves to use tools,'' \textit{Advances in Neural Information Processing Systems}, vol.~36, 2024.

\bibitem{yao2023react}
S.~Yao, J.~Zhao, D.~Yu, N.~Du, I.~Shafran, K.~Narasimhan, and Y.~Cao, ``ReAct: Synergizing reasoning and acting in language models,'' \textit{Proceedings of the International Conference on Learning Representations}, 2023.

\end{thebibliography}
\end{document}